# Crystalline anisotropic magnetoresistance with two-fold and eight-fold symmetry in (In,Fe)As ferromagnetic semiconductor

Pham Nam Hai, Daisuke Sasaki, Le Duc Anh and Masaaki Tanaka

*Department of Electrical Engineering and Information Systems, The University of Tokyo, 7-3-1 Hongo, Bunkyo-ku, Tokyo 113-8656, Japan*

We have investigated the anisotropic magnetoresistance (AMR) of (In,Fe)As ferromagnetic semiconductor (FMS) layers grown on semi-insulating GaAs substrates. In a 10 nm-thick (In,Fe)As layer which is insulating at low temperature, we observed crystalline AMR with two-fold and eight-fold symmetries. In a metallic 100 nm-thick (In,Fe)As layer with higher electron concentration, only two-fold symmetric crystalline AMR was observed. Our results demonstrate the macroscopic ferromagnetism in (In,Fe)As with magnetic anisotropy that depends on the electron concentration. Non-crystalline AMR is also observed in the 100 nm-thick layer, but its magnitude is as small as $10^{-5}$, suggesting that there is no *s-d* scattering near the Fermi level of (In,Fe)As. We propose the origin of the eight-fold symmetric crystalline anisotropy in (In,Fe)As.

Carrier-induced ferromagnetic semiconductors are of great interest because of their possible applications in semiconductor spintronic devices. However, most studies have been concentrated in Mn based p-type FMSs, such as (In,Mn)As[1,2] and (Ga,Mn)As[3,4]. Recently, we have successfully fabricated a new Fe based n-type electron-induced ferromagnetic semiconductor (In,Fe)As.[5] This new material has many advantages over the conventional Mn based FMSs, such as the ability to control ferromagnetism by both Fe and independent carrier doping. In this letter, we investigate the in-plane anisotropic magnetoresistance (AMR) of (In,Fe)As layers with different electron concentrations. In single crystal ferromagnets, AMR can be decomposed into a dominating non-crystalline term and usually weaker crystalline terms which reflect the magnetocrystalline anisotropies. Here, we show that AMR of (In,Fe)As layers are instead dominated by crystalline AMR with the two-fold and eight-fold symmetries, and their magnitudes and signs change with electron concentration. The eight-fold symmetry is the highest order of symmetry ever observed in ferromagnets. We will show that the eight-fold symmetry can be explained by the crystal field in zinc-blende crystal structure. A non-crystalline AMR term is also observed, but its magnitude is as small as $10^{-5}$. The very small non-crystalline AMR term suggests that there is no *s-d* scattering near the Fermi level; this means there is no impurity *d*-state at the Fermi level of (In,Fe)As.

We have prepared two samples for AMR measurements. One is a 10 nm-thick $(In_{0.94},Fe_{0.06})As$ layer (sample A) with an electron concentration of $8\times10^{18}$ cm$^{-3}$. The other is a 100 nm-thick $(In_{0.95},Fe_{0.05})As$ layer (sample B) with an electron concentration of $1.8\times10^{19}$ $cm^{-3}$. Both samples were grown by low-temperature molecular beam epitaxy on semi-insulating GaAs(001) substrates. Electrons in (In,Fe)As are supplied by co-doped Be double-donors. Details of the epitaxial growth and ferromagnetic properties of our (In,Fe)As samples are reported elsewhere.[5] The electron concentration was estimated by Hall effect measurements at room temperature. Figure 1 shows the temperature dependence of the resistivity of sample A and B. While sample A becomes insulating at low temperature, sample B remains metallic even at low temperature. The Curie temperatures are 30 K and 34 K for sample A and B, respectively, which were estimated by the Arrort plots of the magnetic circular dichroism hysteresis curves at a photon energy of 2.6 eV.

From symmetry arguments, in-plane AMR and planar Hall resistance (PHR) of a cubic crystal can be phenomenologically given by [6,7]

$$\Delta\rho_{xx}/\rho_{avr} = C_1\cos(2\phi) + C_{1,c}\cos(4\varphi - 2\phi) + C_2\cos(2\varphi) + C_4\cos(4\varphi), \quad (1a)$$

$$\Delta\rho_{xy}/\rho_{avr} = C_1\sin(2\phi) - C_{1,c}\sin(4\varphi - 2\phi). \quad (1b)$$

Here $\Delta\rho_{xx} = \rho_{xx} - \rho_{avr}$, with $\rho_{avr}$ the averaged longitudinal resistivity, $\phi$ is the angle between the magnetization $M$ and the current $i$, $\varphi$ is the angle between $M$ and the [110] crystal axis. The first term is the non-crystalline AMR. In polycrystalline ferromagnetic metals, this term is dominant[8] and is shown to originate from the $s$-$d$ scattering.[9,10,11] The second term is the crossed non-crystalline / crystalline terms. The third and the fourth terms are crystalline AMR terms, which are related to the symmetry of the crystal, and can be observed only in single crystal samples. The two-fold symmetric term $C_2\cos(2\varphi)$ originally does not exist in ideal cubic crystals, but it is added to describe an additional two-fold symmetry along the [110] direction as actually observed in the case of (Ga,Mn)As[6,7] and (In,Mn)As.[12] Higher order crystalline terms are usually neglected in Eq. (1a).

For AMR measurements of sample A and B, we fabricated Hall bars with a width of 50 μm and current/voltage terminals separated by 200 μm along the [110] and [$\bar{1}$10] directions. $\rho_{xx}$ of the Hall bars were measured by the DC 4-terminal method. A magnetic field of $H = 8.7$ kG was applied in-plane so that the magnetization $M$ are completely aligned along the external field direction. The magnetic field $H$ was rotated around the samples from $\alpha = 0$ to 360°, where $\alpha = \varphi + \pi/4$ is the angle between $H$ and the [100] crystal axis. Inset in Fig. 1 shows the relationships between $i$, $H$, $M$, and the angles $\alpha$, $\varphi$, and $\phi$ in our measurements.

Figures 2(a) and (b) show the $\alpha$ polar plot of $\Delta\rho_{xx}/\rho_{\text{avr}} - \Delta\rho_{xx\text{-min}}/\rho_{\text{avr}}$ for the [110] and [$\bar{1}$10] Hall bars of sample A, measured at 20 K with a current of 10 μA. The squares are experimental data and the solid curves are fitting curves. We clearly observed a dominant two-fold symmetric AMR peaks along the [$\bar{1}$10] direction. There are additional AMR peaks at approximately 0, π/4, π/2, π, 5π/4 and 3π/2, revealing eight-fold symmetry. Note that the positions of these peaks do not depend on the current direction, as no shift of these peaks was observed when we changed the current direction from [110] to [$\bar{1}$10]. Therefore, the AMR of sample A is dominated by the crystalline term rather than the non-crystalline term. We then neglect the non-crystalline related terms and rewrite Eq. (1a) as follows:

$$\Delta\rho_{xx}/\rho_{avr} \sim C_2\cos(2\varphi) + C_4\cos(4\varphi) + C_8\cos(8\varphi). \tag{3}$$

The $C_2$, $C_4$ and $C_8$ coefficients are estimated from the peaks of $\Delta\rho_{xx}/\rho_{avr}$ at $\varphi = 0$, $\pi/4$ and $\pi/2$. The results are $C_2 = -0.038\%$, $C_4 \sim 0\%$, $C_8 = 0.025\%$ for the [110] Hall bar, and $C_2 = -0.052\%$, $C_4 \sim 0\%$, $C_8 = 0.03\%$ for the [$\bar{1}$10] Hall bar. The full fitting curves calculated using these values in Eq. (3) are shown by the solid curves in Fig. 2, and explain reasonably well the experimental results (squares). As far as we know, such eight-fold symmetric crystalline AMR has never been reported in any ferromagnets. The eight-fold symmetry is therefore a unique properties of zinc-blende (In,Fe)As with tetrahedral Fe-As bonding.

Figures 3(a) and (b) show the $\alpha$ polar plot of $\Delta\rho_{xx}/\rho_{avr} - \Delta\rho_{xx\text{-min}}/\rho_{avr}$ for the [110] and [$\bar{1}$10] Hall bars of sample B, measured at 20 K. The current is 1 mA. Sample B shows two-fold symmetric peaks along the [110] direction. Again, these peaks can be attributed to the crystalline AMR components, since their positions do not change when the current is changed from [110] to [$\bar{1}$10] direction. In contrast to sample A, there is no eight-fold symmetric component in sample B. The $C_2$ coefficients are 0.0034% and 0.12% for the [110] and [$\bar{1}$10] Hall bar, respectively. Our results clearly show that the symmetries of crystalline AMR and their signs in (In,Fe)As change with the electron concentration. Such changes of the symmetries of crystalline AMR and their signs have been observed in (Ga,Mn)As: Rushforth *et al.* observed that crystalline AMR of a 25 nm-thick (Ga,Mn)As layer has four-fold symmetry and two-fold symmetry along the [$\bar{1}$10] direction, but that of a 5 nm-thick (Ga,Mn)As has only two-fold symmetry and its peaks changed to the [110] direction.[7]

There are two implications from the above results. Firstly, the existence of crystalline AMR in our (In,Fe)As samples shows that there is intrinsic macroscopic ferromagnetism in our single-crystal (In,Fe)As layers. If these (In,Fe)As layers would be non-magnetic but contained superparamagnetic Fe-related nanoclusters, the magnetoresistance would be dominated by the tunneling magnetoresistance effect between adjacent nanoclusters. In this case, the magnetoresistance should be isotropic, i.e. there should be no dependence on the magnetic field direction. Clearly, this is not the case for (In,Fe)As. Secondly, it has been shown in (Ga,Mn)As that the symmetries of crystalline AMR reflect those of the magnetocrystalline anisotropies.[7] We can prove this relationship strictly using Neumann's principle as follows: The magnetocrystalline anisotropy energy $V(\hat{\alpha})$ and the resistivity tensor $\rho_{ij}(\hat{\alpha})$ can be expanded in a MacLaurin's series:[13]

$$V(\hat{\alpha}) = b_{ij}\alpha_i\alpha_j + b_{ijkl}\alpha_i\alpha_j\alpha_k\alpha_l + ..., \tag{4a}$$

$$\rho_{ij}(\hat{\alpha}) = c_{ij} + c_{kij}\alpha_k + c_{klij}\alpha_k\alpha_l + c_{klmij}\alpha_k\alpha_l\alpha_m + ..., \tag{4b}$$

here $\hat{\alpha}$ is the unit vector of the magnetization direction in the crystal frame of reference. The subscripts $i, j,... \in (1,2,3) \equiv (x,y,z)$ correspond to the three orthogonal coordinate axes of the crystal under consideration. According to the Neumann's principle, the coefficient tensors $b_{ij}, b_{ijkl}, c_{ij}, c_{kij},\ldots$ should be unchanged under symmetry operations using the generating matrix $S$ of the considered crystal: $b_{kl} = |S|^2 S_{ks} S_{lt} b_{st}$, $b_{klst} = |S|^4 S_{ko} S_{lm} S_{sp} S_{tq} b_{ompq}$, $c_{kl} = |S|^2 S_{ks} S_{lt} c_{st}$, $c_{kls} = |S|^3 S_{ko} S_{lm} S_{sp} c_{omp}$ ... The longitudinal resistivity when the current is in the $\hat{\beta}$ direction is given by

$$\rho(\hat{\alpha},\hat{\beta}) = \beta_i \rho_{ij}(\hat{\alpha}) \beta_j = c_{ij}\beta_i\beta_j + c_{klij}\alpha_k\alpha_l\beta_i\beta_j + c_{klmnij}\alpha_k\alpha_l\alpha_m\alpha_n\beta_i\beta_j + \ldots \quad (5)$$

In Eq. (5), only the symmetric part of the $\rho_{ij}$ tensor contributes to the longitudinal resistivity. To calculate the crystalline AMR component, we integrate Eq. (5) for all $\hat{\beta}$ directions, which results in

$$\rho_{crys}(\hat{\alpha}) = \frac{1}{3}(c_{ii} + c_{klii}\alpha_k\alpha_l + c_{klmnii}\alpha_k\alpha_l\alpha_m\alpha_n + \ldots) \equiv d_0 + d_{kl}\alpha_k\alpha_l + d_{klmn}\alpha_k\alpha_l\alpha_m\alpha_n + \ldots,$$

$$d_{kl} \equiv (c_{kl11} + c_{kl22} + c_{kl33})/3 = (1/3)|S|^4 S_{ko} S_{lm} (S_{11}^2 c_{om11} + S_{12}^2 c_{om22} + S_{13}^2 c_{om33} + \sum_{p \neq q} S_{1p} S_{1q} c_{ompq}$$
$$+ S_{21}^2 c_{om11} + S_{22}^2 c_{om22} + S_{23}^2 c_{om33} + \sum_{p \neq q} S_{2p} S_{2q} c_{ompq} + S_{31}^2 c_{om11} + S_{32}^2 c_{om22} + S_{33}^2 c_{om33} + \sum_{p \neq q} S_{3p} S_{3p} c_{ompq})$$

Since $S_{kp} S_{kq} = \delta_{pq}$ and $|S|^2 = 1$ for a frame of reference with orthogonal coordinate axes,[13] we can see that $d_{kl}$ is also unchanged under the same symmetry operations:

$$d_{kl} = (1/3)|S|^2 S_{ko} S_{lm} (c_{om11} + c_{om22} + c_{om33}) = |S|^2 S_{ko} S_{lm} d_{om}.$$

In this way, we proved for the general case that the crystalline AMR has the same symmetries as the magnetocrystalline anisotropy energy (possible microscopic mechanism of this similarity will be discussed later). Therefore, our AMR results reveal that there are two-fold and eight-fold symmetries in magnetocrystalline anisotropy in the 10 nm-thick (In,Fe)As layer, while there is only two-fold symmetric magnetocrystalline anisotropy in the 100 nm-thick layer.

Next, we measured the planar Hall effect to detect the non-crystalline components. According to Eq. (1b), there is no crystalline component in the planar Hall resistivity, thus it is much easier to detect the non-crystalline components in PHR, if any. Figures 4(a) and (b) show the $\phi$ polar plot of $\Delta\rho_{xy}/\rho_{avr}$ - $\Delta\rho_{xy\text{-min}}/\rho_{avr}$ for the [110] and [$\bar{1}$10] Hall bars of sample B, measured at 20 K, respectively. The current is 5 mA for the [110] Hall bar, and 1 mA for the [$\bar{1}$10] Hall bar. It can be seen that the PHR depend only on $\phi$ (angle between **M** and **i**), and not on $\alpha$ (angle between **M** and the [100] crystal axis) as in the case of longitudinal AMR. The black curves are fitting curves using $C_1 = -1.8\times 10^{-5}$, $C_{1,c} = 0$ for the [110] Hall bar and $C_1 = -2.8\times 10^{-5}$, $C_{1,c} = 0$ for the [$\bar{1}$10] Hall bar in Eq. (1b), respectively. For sample A, we can not observe any non-crystalline terms.

Let us discuss on the microscopic mechanism of our results and their indications on the electronic structure of (In,Fe)As. The scattering rate of an electron from the initial state $\psi_i$ to the final state $\psi_f$ by a scattering potential $V$ is given by $\Gamma_{i\to f} = \frac{2\pi}{\hbar}|\langle\psi_f|V|\psi_i\rangle|^2 \delta(E_i - E_f)$. When there are both d-state wave functions $\psi_d$ and s-state wave functions $\psi_s$ at the Fermi level, scattering from the $\psi_s$ to $\psi_d$ that is mixed under the effect of spin-orbit interaction will depend on the direction between magnetization and current, which results in large non-crystalline AMR even when the scattering potential $V$ is isotropic. This s-d scattering is the well-know mechanism of non-crystalline AMR in ferromagnetic metals.[9,10,11] In our (In,Fe)As, the dominant scattering mechanism is neutral impurity scattering by neutral Fe atoms, which is also isotropic.[5] The very small non-crystalline components in (In,Fe)As indicate that there seem to be no or very little *s-d* scattering in electron transport near the Fermi level of (In,Fe)As. This fact is consistent with our observation that electron carriers (and Fermi level) in (In,Fe)As reside in the conduction band rather than in the *d*-state impurity band.[5] On the other hand, the crystalline AMR in (In,Fe)As can reach the order of 0.1%, which is comparable with those of (Ga,Mn)As.[6,7] We consider a scattering rate in (In,Fe)As in the following form: $\Gamma_{s \to s'} = \frac{2\pi}{\hbar} |\langle \psi_{s'} | J_{sd} (\mathbf{s} \cdot \mathbf{S}) | \psi_s \rangle|^2 \delta(E_s - E_{s'}) \delta(\mathbf{r} - \mathbf{R})$. Here (*s*,*S*) and (*r*,*R*) are the spin and position of a conduction *s*-electron and a localized *d*-electron, respectively. $J_{sd}(\mathbf{s} \cdot \mathbf{S})$ is the *s-d* exchange interaction between the conduction and localized electrons. Because $\psi_s$ is isotropic, the symmetries of the scattering rate are the same as those of magnetization $\mathbf{M} = \sum \mathbf{S}$. The scattering process in this form can be a possible microscopic origin for the theorem we proved above using the phenomenological symmetry approach.

Next, we propose a mechanism that can explain the eight-fold symmetry of the crystalline AMR and magnetization in (In,Fe)As. Due to the strong spin-orbit interaction $\lambda \mathbf{L} \cdot \mathbf{S}$ in (In,Fe)As, localized spins *S* are coupled to their orbital wave functions, which are coupled to the lattice by crystal fields. Figures 5(a) and (b) show the plan view of nearest In/As atoms around an Fe atom along each direction in the *x-y* plane, respectively, and the orbital wave function $d_{xy}$ when $S$ is along the <100> and <110> direction. Suppose that the effective charge of an In atom (including screening by the covalent electrons) is $+\varepsilon e$, the Coulomb potential $E_{\text{In-}d}$ of a $d$-electron from this charge is four-fold symmetric with *maximums* along the <100> and <010> directions. On the other hand, the As atoms can have an effective negative charge of $-\gamma e$, and also form a four-fold symmetric potential $E_{\text{As-}d}$ with *minimums* along the <100> and <010> directions in the $x$-$y$ plane. The crystal field energies $E_{\text{In-}d}$ and $E_{\text{As-}d}$ due to In/As atoms can be expanded to Fourier series $E_{\text{In-}d} = A_4\cos(4\alpha) + A_8\cos(8\alpha) +\ldots$ and $E_{\text{As-}d} = -B_4\cos(4\alpha) - B_8\cos(8\alpha) -\ldots$. The total crystal field energy $E_{\text{total-}d} = (A_4-B_4)\cos(4\alpha) + (A_8-B_8)\cos(8\alpha) + \ldots$ can have a dominant eight-fold symmetry when $A_4 = B_4$. It should be noted that this is caused by the zinc-blende lattice structure of (In,Fe)As. In other cubic crystals, such as the fcc Fe crystal, such a high order of symmetry does not appear. The disappearance of eight-fold symmetry in sample B with higher electron concentration may be explained by the increasing screening effect of free electrons that virtually reduce $+\varepsilon e$ and $-\gamma e$ to zero, i.e., no crystal field. However, the two-fold symmetry along the [110] or [$\bar{1}$10] directions can not be explained by the crystal structure of (In,Fe)As. Such an uniaxial anisotropy has also been observed in (Ga,Mn)As [6,7] and (In,Mn)As[12], but the physical origin is still unclear. The switching of the symmetry axis from [$\bar{1}$10] to [110] directions when the carrier concentration increases has also been observed in (Ga,Mn)As.[14]

In conclusion, we have investigated AMR in (In,Fe)As ferromagnetic semiconductor layers. The longitudinal AMR in (In,Fe)As is dominated by two-fold and eight-fold symmetric crystalline components, whose magnitudes and signs depend on the electron concentration. These results indicate the intrinsic macroscopic ferromagnetism in (In,Fe)As, and reveal two-fold and eight-fold symmetry of magnetocrystalline anisotropy. We also observed a very small non-crystalline component, which is of the order of $10^{-5}$. The very small non-crystalline component suggests that there is no *s-d* scattering near the Fermi level of (In,Fe)As, which is consistent with our previous observation that electron carriers resides in the conduction-band of (In,Fe)As rather than in the *d*-state impurity band. We also show that the eight-fold symmetry can be explained by the crystal field of the zinc-blende structure acting on localized *d*-electrons.

**Acknowledgments**

This work was partly supported by Grant-in-Aids for Scientific Research, particularly the Grant-in-Aid for Specially Promoted Research, the Special Coordination Programs for Promoting Science and Technology, the FIRST Program of JSPS, and the Global COE program (C04).

# References

[1] S. Koshihara *et al.*, Phys. Rev. Lett. **78**, 4617 (1997).

[2] H. Ohno *et al.*, Nature **408**, 944 (2000).

[3] A. M. Nazmul, S. Kobayashi, S. Sugahara and M. Tanaka, Jpn. J. Appl. Phys. **43**, L233 (2004).

[4] D. Chiba, F. Matsukura, H. Ohno, Appl. Phys. Lett. **89**, 162505 (2006).

[5] P. N. Hai, L. D. Anh and M. Tanaka, arXiv:1106.0561v3 [cond-mat.mtrl-sci].

[6] A. W. Rushforth *et al.*, Phys. Rev. Lett. **99**, 147207 (2007).

[7] A. W. Rushforth *et al.*, J. Magn. Magn. Mat. **321**, 1001 (2009).

[8] T. McGuire and R. Potter, IEEE Trans. Magn. **11**, 1018 (1975).

[9] J. Smit, Physica **6**, 612 (1951).

[10] R. I. Potter, Phys. Rev. B **10**, 4626 (1974).

[11] O. Jaoul, I.A. Campbell and A. Fert, J. Magn. Magn. Mat. **5**, 23 (1977).

[12] X. Liu *et al.*, Appl. Phys. Lett. **86**, 112512 (2005).

[13] R. R. Birss, *Symmetry and Magnetism* (North-Holland, Amsterdam, 1966).

[14] M. Sawicki *et al.*, Phys. Rev. B **71**, 121302 (2005).

**Figure captions**

**Fig. 1.** (color online) Temperature dependence of resistivity of sample A (solid curve) and sample B (dashed curve). Inset shows relationships between the current $i$, the magnetic field $H$, the magnetization $M$, and angles $\alpha$, $\varphi$ and $\phi$ in our measurements. $M$ is parallel to $H$ because of the high field (8.7 kOe). $\alpha$ is the angle between $M$ and the [100] crystal axis, $\varphi$ is the angle between $M$ and the [110] crystal axis, and $\phi$ is the angle between $M$ and $i$.

**Fig. 2.** (color online) Longitudinal anisotropic magnetoresistance data (squares) for **(a)** the [110] Hall bar ($i$ = 10 μA // [110]), and **(b)** the [$\bar{1}$10] Hall bar ($i$ = 10 μA // [$\bar{1}$10]) of sample A as a function of $\alpha$, measured at 20 K under a magnetic field of 8.7 kG. Solid curves are fitting curves using $C_2$ = -0.038%, $C_4$ = 0%, $C_8$ = 0.025% for the [110] Hall bar, and $C_2$ = -0.052%, $C_4$ = 0%, $C_8$ = 0.03% for the [$\bar{1}$10] Hall bar in Eq. (3), respectively.

**Fig. 3.** (color online) Longitudinal anisotropic magnetoresistance data (squares) for **(a)** the [110] Hall bar ($i$ = 1 mA // [110]), and **(b)** the [$\bar{1}$10] Hall bar ($i$ = 1 mA // [$\bar{1}$10]) of sample B as a function of $\alpha$, measured at 20 K under a magnetic field of 8.7 kG. Solid curves are fitting curves using $C_2$ = 0.0034%, $C_4$ = 0%, $C_8$ = 0% for the [110] Hall bar, and $C_2$ = 0.012%, $C_4$ = 0%, $C_8$ = 0% for the [$\bar{1}$10] Hall bar in Eq. (3), respectively.

**Fig. 4.** (color online) Planar Hall resistance data (squares) for **(a)** the [110] Hall bar ($i$ = 5 mA // [110]), and **(b)** the [$\bar{1}$10] Hall bar ($i$ = 1 mA // [$\bar{1}$10]) of sample B as a function of $\phi$, measured at 20 K under a magnetic field of 8.7 kG. Solid curves are fitting curves using $C_1$ = -1.8×10$^{-5}$, $C_{1,c}$ = 0 for the [110] Hall bar, and $C_1$ = -2.8×10$^{-5}$, $C_{1,c}$ = 0 for the [$\bar{1}$10] Hall bar in Eq. (1b), respectively.

**Fig. 5.** (color) Plain view of nearest **(a)** In atoms and **(b)** As atoms around an Fe atom. In atoms have an effective charge of +$\varepsilon e$, while As atoms have an effective charge of -$\gamma e$, seen from a $d$-electron. A $d$-electron with the $d_{xy}$ wave function will be closer to its nearest In/As atoms when its wave function is aligned along the <110> direction (dashed red curve) rather than the <100> direction (solid red curve).

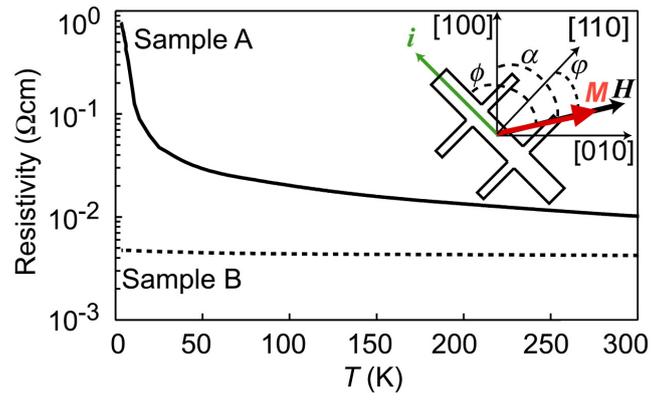

Fig. 1. Hai *et al.*

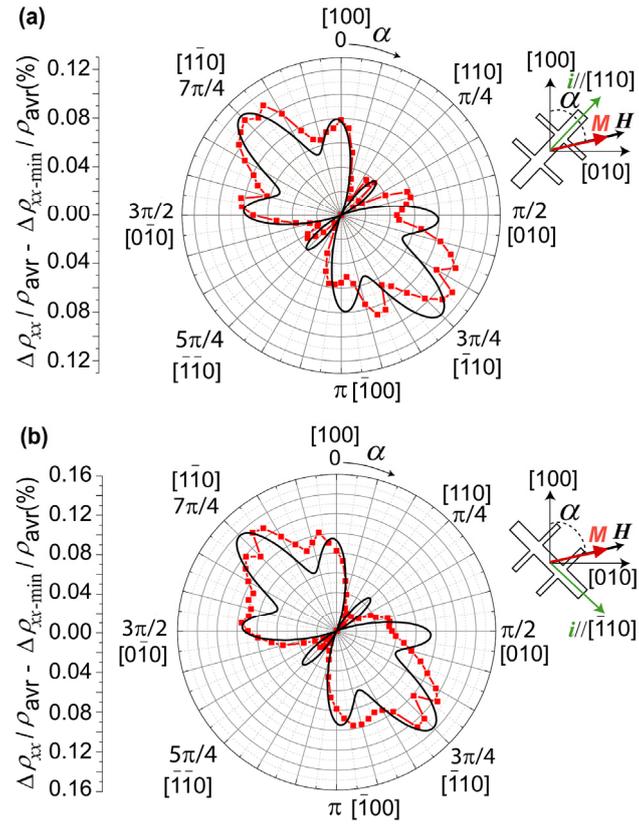

Fig. 2. Hai *et al.*

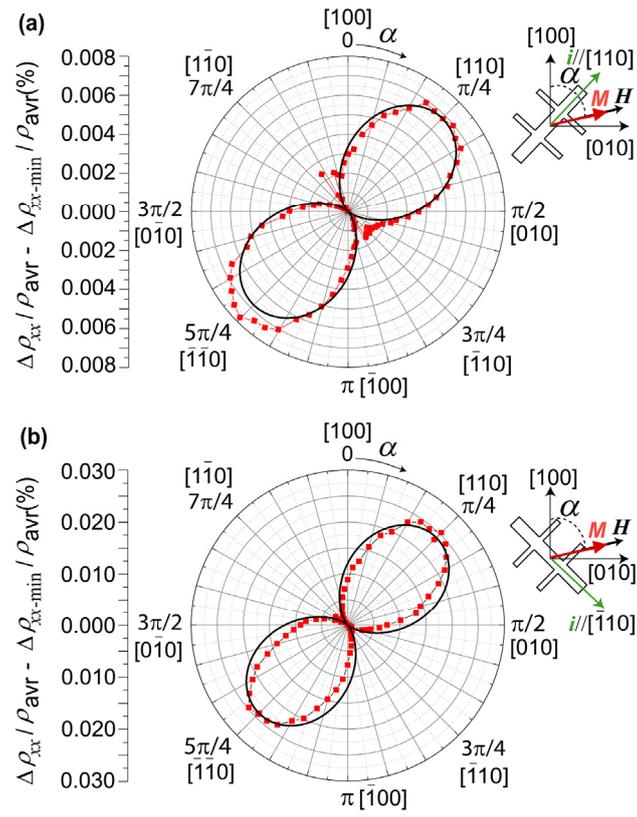

Fig. 3. Hai *et al.*

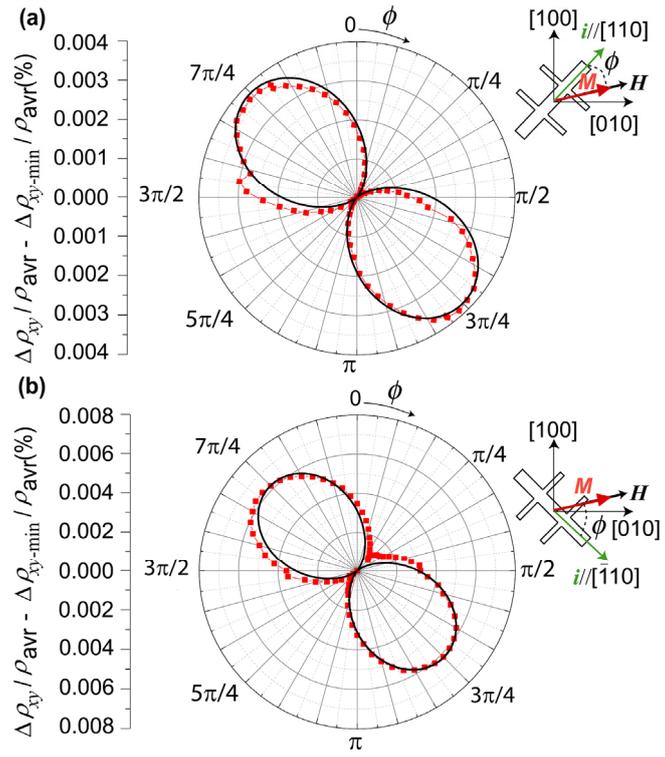

Fig. 4. Hai *et al.*

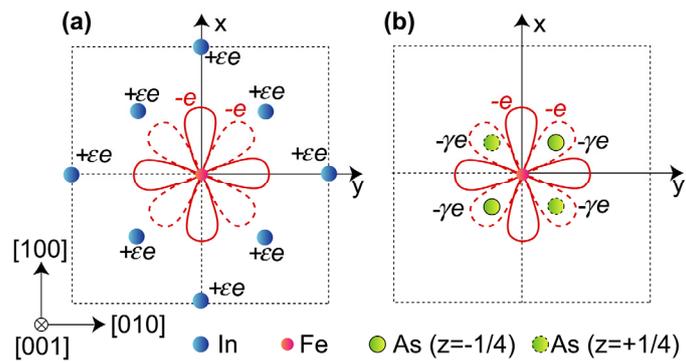

Fig. 5. Hai *et al.*